\documentstyle[12pt]{article}

\font\twlgot =eufm10 scaled \magstep1 \font\egtgot =eufm8
\font\sevgot =eufm7 \font\twlmsb =msbm10 scaled \magstep1
\font\egtmsb =msbm8 \font\sevmsb =msbm7

\newfam\gotfam
\def\pgot{\fam\gotfam\twlgot}
\textfont\gotfam\twlgot \scriptfont\gotfam\egtgot
\scriptscriptfont\gotfam\sevgot
\def\got{\protect\pgot}
\newfam\msbfam
\textfont\msbfam\twlmsb \scriptfont\msbfam\egtmsb
\scriptscriptfont\msbfam\sevmsb
\def\Bbb{\protect\pBbb}
\def\pBbb{\relax\ifmmode\expandafter\Bb\else\typeout{You cann't use
Bbb in text mode}\fi}
\def\Bb #1{{\fam\msbfam\relax#1}}

\newcommand{\gQ}{{\got T}}

\newcommand{\gA}{{\got A}}

\newcommand{\gd}{{\got d}}

\newcommand{\gS}{{\got S}}

\def\thebibliography#1{\bigskip\section*{}\bigskip\list
{$^{\arabic{enumi}}$}{\settowidth\labelwidth{#1}\leftmargin\labelwidth
\advance\leftmargin\labelsep
\usecounter{enumi}}
\def\newblock{\hskip .11em plus .33em minus .07em}
\sloppy\clubpenalty4000\widowpenalty4000 \sfcode`\.=1000\relax}

\def\op#1{\mathop{\fam0 #1}\limits}

\newcommand{\Ker}{{\rm Ker\,}}
\newcommand{\im}{{\rm Im\,}}
\newcommand{\nm}[1]{|{#1}|}
\newcommand{\beq}{\begin{equation}}
\newcommand{\eeq}{\end{equation}}
\newcommand{\ben}{\begin{eqnarray}}
\newcommand{\een}{\end{eqnarray}}
\newcommand{\be}{\begin{eqnarray*}}
\newcommand{\ee}{\end{eqnarray*}}
\newcommand{\bea}{\begin{eqalph}}
\newcommand{\eea}{\end{eqalph}}
\newcommand{\cA}{{\cal A}}

\newcommand{\cP}{{\cal P}}

\newcommand{\cL}{{\cal L}}

\newcommand{\cV}{{\cal V}}
\newcommand{\cE}{{\cal E}}

\newcommand{\cS}{{\cal S}}
\newcommand{\cC}{{\cal C}}
\newcommand{\cO}{{\cal O}}

\newcommand{\cK}{{\cal K}}
\newcommand{\bL}{{\bf L}}

\newcommand{\bE}{{\bf E}}

\newcommand{\al}{\alpha}
\newcommand{\vr}{\varrho}
\newcommand{\bt}{\beta}
\newcommand{\dl}{\delta}
\newcommand{\la}{\lambda}
\newcommand{\La}{\Lambda}
\newcommand{\f}{\phi}
\newcommand{\om}{\omega}
\newcommand{\Om}{\Omega}
\newcommand{\m}{\mu}

\newcommand{\G}{\Gamma}
\newcommand{\th}{\theta}

\newcommand{\vt}{\vartheta}

\newcommand{\up}{\upsilon}

\newcommand{\di}{{\rm dim\,}}

\newcommand{\e}{\epsilon}

\newcommand{\si}{\sigma}
\newcommand{\Si}{\Sigma}
\newcommand{\w}{\wedge}

\newcommand{\ol}{\overline}
\newcommand{\dr}{\partial}
\newcommand{\ar}{\op\longrightarrow}
\newcommand{\llr}{\op\longleftarrow}
\newcommand{\lto}{\leftarrow}
\newcommand{\ot}{\otimes}

\newcommand{\rdr}{\stackrel{\leftarrow}{\dr}{}}

\let\ssection=\section
\renewcommand{\section}{\setcounter{equation}{0}\ssection}

\newcounter{eqalph}
\newcounter{equationa}
\newcounter{remark}
\newcounter{example}
\newcounter{theorem}
\newcounter{proposition}
\newcounter{lemma}
\newcounter{corollary}
\newcounter{definition}
\newenvironment{eqalph}{\stepcounter{equation}
\setcounter{equationa}{\value{equation}} \setcounter{equation}{0}

\begin{eqnarray}}{\end{eqnarray}\setcounter{equation}{\value{equationa}}}

\def\theremark{\arabic{remark}}

\def\thetheorem{\arabic{theorem}}

\newenvironment{proof}{
{\it Proof:}}{}
\newenvironment{rem}{\refstepcounter{remark}{\it
Remark \theremark:}}{}

\newenvironment{theo}{\refstepcounter{theorem}
{\bf Theorem \thetheorem:}}{}
\newenvironment{prop}{\refstepcounter{theorem}
{\bf Proposition \thetheorem:}}{}
\newenvironment{lem}{\refstepcounter{theorem}
{\bf Lemma \thetheorem:}}{}

\newenvironment{defi}{\refstepcounter{theorem}
{\bf Definition \thetheorem:}}{}

\newcommand{\mar}[1]{}

\begin{document}
\hbox{}

{\parindent=0pt

{\large\bf The antifield Koszul--Tate complex of reducible Noether
identities}
\bigskip

{\sc D.Bashkirov}\footnote{Electronic mail: bashkir@phys.msu.ru}

{\sl Department of Theoretical Physics, Moscow State University,
117234 Moscow, Russia}

\medskip

{\sc G.Giachetta}\footnote{Electronic mail:
giovanni.giachetta@unicam.it}

{\sl Department of Mathematics and Informatics, University of
Camerino, 62032 Camerino (MC), Italy}

\medskip

{\sc L.Mangiarotti}\footnote{Electronic mail:
luigi.mangiarotti@unicam.it}

{\sl Department of Mathematics and Informatics, University of
Camerino, 62032 Camerino (MC), Italy}

\medskip

{\sc G. Sardanashvily}\footnote{Electronic mail:
sard@grav.phys.msu.su}

{\sl Department of Theoretical Physics, Moscow State University,
117234 Moscow, Russia}

\bigskip

A generic degenerate Lagrangian system of even and odd fields is
examined in algebraic terms of the Grassmann-graded variational
bicomplex. Its Euler--Lagrange operator obeys Noether identities
which need not be independent, but satisfy first-stage Noether
identities, and so on. We show that, if a certain necessary and
sufficient condition holds, one can associate to a degenerate
Lagrangian system the exact Koszul--Tate complex with the boundary
operator whose nilpotency condition restarts all its Noether and
higher-stage Noether identities. This complex provides a
sufficient analysis of the degeneracy of a Lagrangian system for
the purpose of its BV quantization.

}

\bigskip
\bigskip

\noindent {\bf I. INTRODUCTION}
\bigskip

As well-known, quantization of a Lagrangian field system
essentially depends on the analysis of its degeneracy. One says
that a Lagrangian system is degenerate if its  Euler--Lagrange
operator obeys non-trivial Noether identities. They need not be
independent, but satisfy the first-stage Noether identities, which
in turn are subject to the second-stage ones, and so on. The
hierarchy of reducible Noether identities characterizes the
degeneracy of a Lagrangian system in full. Noether's second
theorem states the relation between the Noether identities and the
gauge symmetries of a Lagrangian system.$^{1,2}$ If Noether
identities and gauge symmetries are finitely generated, they are
parameterized by the modules of antifields and ghosts,
respectively. An original Lagrangian is extended to these
antifields and ghosts in order to satisfy the so-called master
equation. This extended Lagrangian is the starting point of the
Batalin-Vilkovisky (BV) quantization of a degenerate Lagrangian
field system.$^{3,4}$

Let us note that the notion of a reducible Noether identity has
come from that of a reducible constraint. Their Koszul--Tate
complex has been invented by analogy with that of constraints$^5$
under a rather restrictive regularity condition that field
equations as well as Noether identities of arbitrary stage can be
locally separated into the independent and dependent ones.$^{6,7}$
This condition also comes from the case of a constraint locally
given by a finite number of functions to which the inverse mapping
theorem can be applied. In contrast with constraints, Noether and
higher-stage Noether identities are differential operators. They
are locally given by a set of functions and their jet
prolongations on an infinite order jet manifold. Since the latter
is a Fr\'echet, but not Banach manifold, the inverse mapping
theorem fails to be valid. Here, we follow the general definition
of Noether identities of differential operators.$^8$ This
definition reproduces that in Refs. [1,2] if Noether identities
are finitely generated. Their Koszul--Tate complex is constructed
iff a certain homology regularity condition holds.

Our goal is the following. Bearing in mind BV quantization, we
address a generic Lagrangian systems of even and odd fields on an
arbitrary smooth manifold $X$ ($\di X=n$). It is algebraically
described in terms of a certain bigraded differential algebra
(henceforth BGDA) $\cS^*_\infty[F;Y]$ which is split into the
Grassmann-graded variational bicomplex, generalizing the
variational bicomplex on fiber bundles (Section II). If a fiber
bundle $Y\to X$ of even fields is affine, this algebra has been
defined as the product of graded algebras of odd and even
fields.$^{2,9}$ Here, its definition is generalized to an
arbitrary fiber bundle $Y\to X$. In this case, elements of
$\cS^*_\infty[F;Y]$ are Grassmann-graded differential forms on the
infinite order jet manifold $J^\infty Y$ of sections of $Y\to X$,
but not on $X$. Let $L\in \cS^{0,n}_\infty[F;Y]$ be a Lagrangian
and $\dl L \in \cS^{1,n}_\infty[F;Y]$ its Euler--Lagrange
operator. We associate to $\dl L$ the chain complex (\ref{v42})
whose boundaries vanish on-shell, i.e., on $\Ker\dl L$
(Proposition \ref{v120}). It is a complex of a certain
$C^\infty(X)$-module $\cP^{0,n}_\infty[\ol Y^*;F;Y;\ol F^*]$  of
Grassmann-graded densities on the infinite order jet manifold
$J^\infty Y$. For our purpose, this complex can be replaced with
the short zero-exact complex $\cP^{0,n}_\infty[\ol Y^*;F;Y;\ol
F^*]_{\leq 2}$ (\ref{v042}).

\begin{rem}
If there is no danger of confusion, elements of homology are
identified to its representatives. A chain complex is called
$r$-exact if its homology of $k\leq r$ is trivial.
\end{rem}

The Noether identities of the Euler--Lagrange operator $\dl L$ are
defined as nontrivial elements of the first homology $H_1(\ol\dl)$
of the complex (\ref{v042}) (Definition \ref{v121}). Let this
homology be finitely generated  by a projective graded
$C^\infty(X)$-module of finite rank. In accordance with the
Serre--Swan theorem generalized to graded manifolds (Theorem
\ref{v0}), one can introduce the corresponding module of
antifields and extend  the complex (\ref{v042}) to the one-exact
complex $\cP^{0,n}_\infty[\ol E^*\ol Y^*;F;Y;\ol F^*\ol V^*]_{\leq
3}$ (\ref{v66}) with the boundary operator $\dl_0$ (\ref{v204})
whose nilpotency conditions are equivalent to the above-mentioned
Noether identities (Proposition \ref{v137}). First-stage Noether
identities are defined as two-cycles of this complex. They are
trivial if two-cycles are boundaries, but the converse need not be
true. Trivial first-stage Noether identities are boundaries iff a
certain homology condition (called the two-homology regularity
condition) holds (Proposition \ref{v134}). In this case, the
first-stage Noether identities are identified to nontrivial
elements of the second homology of the complex (\ref{v66}). If
this homology is finitely generated, the complex (\ref{v66}) is
extended to the two-exact complex $\cP^{0,n}_\infty[\ol E^*_1\ol
E^*\ol Y^*;F;Y;\ol F^*\ol V^*\ol V^*_1]_{\leq 4}$ (\ref{v87}) with
the boundary operator $\dl_1$ (\ref{v205}) whose nilpotency
conditions are equivalent to the Noether and first-stage Noether
identities (Proposition \ref{v139}). If the third homology of this
complex is not trivial, the second-stage Noether identities exist,
and so on. Iterating the arguments, we come to the following.

Let we have the $(N+1)$-exact complex $\cP^{0,n}_\infty\{N\}_{\leq
N+3}$ (\ref{v94}) such that: (i) the nilpotency conditions of its
boundary operator $\dl_N$ (\ref{v92}) reproduce Noether and
$k$-stage Noether identities for $k\leq N$, (ii) the
$(N+1)$-homology regularity condition holds. This condition states
that any $\dl_{k<N-1}$-cycle $\f\in \cP_\infty^{0,n}\{k\}_{k+3}$
is a $\dl_{k+1}$-boundary (Definition \ref{v155}). Then the
$(N+1)$-stage Noether identities are defined as $(N+2)$-cycles of
this complex. They are trivial if cycles are boundaries, while the
converse is true iff the $(N+2)$-homology regularity condition is
satisfied. In this case, $(N+1)$-stage Noether identities are
identified to nontrivial elements of the $(N+2)$-homology of the
complex (\ref{v94}) (item (i) of Theorem \ref{v163}). Let this
homology is finitely generated. By means of antifields, this
complex is extended to the $(N+2)$-exact complex
$\cP_\infty\{N+1\}_{\leq N+4}$ (\ref{v171}) with the boundary
operator $\dl_{N+1}$ (\ref{v170}) whose nilpotency restarts all
the Noether identities up to stage $(N+1)$ (item (ii) of Theorem
\ref{v163}).

This iteration procedure results in the exact Koszul--Tate complex
of antifields with the boundary operator whose nilpotency
conditions reproduce all Noether and higher Noether identities
characterizing the degeneracy of a differential operator $\dl L$.

In Section V, we address the particular variant of topological BF
theory with the Lagrangian (\ref{v182}) for a scalar $A$ and
$(n-1)$-form $B$ as an example of a reducible degenerate
Lagrangian system$^1$ where the homology regularity condition is
verified (Lemma \ref{v220}), Noether and $k$-stage Noether
identities are proved to be finitely generated, and its
Koszul--Tate complex (\ref{v224}) is constructed.

\begin{rem} \label{v58} \mar{v58}
Throughout the paper, smooth manifolds are assumed to be real,
finite-dimensional, Hausdorff, second-countable (consequently,
paracompact) and connected. By a Grassmann algebra over a ring
$\cK$ is meant a $\Bbb Z_2$-graded exterior algebra of some
$\cK$-module. We restrict our consideration to graded manifolds
$(Z,\gA)$ with structure sheaves $\gA$ of Grassmann algebras of
finite rank.$^{10,11}$ The symbols $|.|$ and $[.]$ stand for the
form degree and Grassmann parity, respectively. We denote by
$\La$, $\Si$, $\Xi$, $\Om$ the symmetric multi-indices, e.g.,
$\La=(\la_1...\la_k)$, $\la+\La=(\la\la_1...\la_k)$. Summation
over a multi-index $\La=(\la_1...\la_k)$ throughout means separate
summation over each its index $\la_i$.
\end{rem}

\bigskip
\bigskip

\noindent {\bf II. GRASSMANN-GRADED LAGRANGIAN SYSTEMS}
\bigskip

Let $Y\to X$ be a fiber bundle and $J^rY$ the jet manifolds of its
sections. They form the inverse system
\mar{5.10}\beq
X\op\longleftarrow^\pi Y\op\longleftarrow^{\pi^1_0} J^1Y
\longleftarrow \cdots J^{r-1}Y \op\longleftarrow^{\pi^r_{r-1}}
J^rY\longleftarrow\cdots, \label{5.10}
\eeq
where $\pi^r_{r-1}$  are affine bundles, and $r=0$ conventionally
stands for $Y$. Its projective limit $(J^\infty
Y;\pi^\infty_r:J^\infty Y\to J^rY)$ is a paracompact Fr\'echet
manifold.  A bundle atlas $\{(U_Y;x^\la,y^i)\}$ of $Y\to X$
induces the coordinate atlas
\mar{jet1}\ben
&& \{((\pi^\infty_0)^{-1}(U_Y); x^\la, y^i_\La)\}, \qquad
{y'}^i_{\la+\La}=\frac{\dr x^\m}{\dr x'^\la}d_\m y'^i_\La, \qquad
0\leq|\La|, \label{jet1} \\
&& d_\la = \dr_\la + \op\sum_{0\leq|\La|} y^i_{\la+\La}\dr_i^\La,
\qquad d_\La=d_{\la_1}\circ\cdots\circ d_{\la_k}, \nonumber
\een
of $J^\infty Y$, where $d_\la$ are total derivatives. We further
assume that the cover $\{\pi(U_Y)\}$ of $X$ is also the cover of
atlases of all fiber bundles over $X$ in question. The inverse
system (\ref{5.10}) yields the direct system
\mar{5.7}\beq
\cO^*X\op\longrightarrow^{\pi^*} \cO^*Y
\op\longrightarrow^{\pi^1_0{}^*} \cO_1^*Y \ar\cdots \cO^*_{r-1}Y
\op\longrightarrow^{\pi^r_{r-1}{}^*}
 \cO_r^*Y \longrightarrow\cdots  \label{5.7}
\eeq
of algebras $\cO_r^*Y$ of exterior forms on jet manifolds $J^rY$
with respect to the pull-back monomorphisms $\pi^r_{r-1}{}^*$. Its
direct limit is the graded differential algebra (henceforth GDA)
$\cO_\infty^*Y$ of all exterior forms on finite order jet
manifolds modulo the pull-back identification.

Let us extend the GDA $\cO_\infty^* Y$ to graded forms on graded
manifolds whose bodies are jet manifolds $J^rY$ of $Y$.$^{2,9}$
Note that there are different approaches to treat odd fields on a
smooth manifold $X$, but the following variant of the Serre--Swan
theorem motivates us to describe them in terms of graded manifolds
whose body is $X$.

\begin{theo} \label{v0} \mar{v0}
Let $Z$ be a smooth manifold. A Grassmann algebra $\cA$ over the
ring $C^\infty(Z)$ of smooth real functions on $Z$ is isomorphic
to the Grassmann algebra of graded functions on a graded manifold
with a body $Z$ iff it is the exterior algebra of some projective
$C^\infty(Z)$-module of finite rank.
\end{theo}

\begin{proof} The proof follows at once from the Batchelor theorem$^{10}$
and the Serre-Swan theorem generalized to an arbitrary smooth
manifold.$^{11,12}$ The Batchelor theorem states that any graded
manifold $(Z,\gA)$ with a body $Z$ is isomorphic to the one
$(Z,\gA_Q)$ with the structure sheaf $\gA_Q$ of germs of sections
of the exterior bundle
\be
\w Q^*=\Bbb R\op\oplus_Z Q^*\op\oplus_Z\op\w^2
Q^*\op\oplus_Z\cdots,
\ee
where $Q^*$ is the dual of some vector bundle $Q\to Z$. Let us
call $(Z,\gA_Q)$ the simple graded manifold with the structure
vector bundle $Q$. Its ring $\cA_Q$ of graded functions (sections
of $\gA_Q$) is the $\Bbb Z_2$-graded exterior algebra of the
$C^\infty(Z)$-module of sections of $\w Q^*\to Z$. By virtue of
the Serre--Swan theorem, a $C^\infty(Z)$-module is isomorphic to
the module of sections of a smooth vector bundle over $Z$ iff it
is a projective module of finite rank.
\end{proof}

In field models, Batchelor's isomorphism is usually fixed from the
beginning. Therefore,  we further consider simple graded manifolds
$(Z,\gA_Q)$. One associates to $(Z,\gA_Q)$ the following BGDA
$\cS^*[Q;Z]$.$^{10}$  Let $\gd\gA_Q$ be the sheaf  of graded
derivations of $\gA_Q$. Its global sections  make up the real Lie
superalgebra $\gd\cA_Q$ of graded derivations of the $\Bbb R$-ring
$\cA_Q$. Then the Chevalley--Eilenberg complex of $\gd\cA_Q$ with
coefficients in $\cA_Q$ can be constructed.$^{13}$ Its subcomplex
$\cS^*[Q;Z]$ of $\cA_Q$-linear morphisms is the Grassmann-graded
Chevalley--Eilenberg differential calculus
\be
0\to \Bbb R\to \cA_Q \ar^d \cS^1[Q;Z]\ar^d\cdots
\cS^k[Q;Z]\ar^d\cdots
\ee
over a $\Bbb Z_2$-graded commutative $\Bbb R$-ring $\cA_Q$. The
graded exterior product $\w$ and Chevalley--Eilenberg coboundary
operator $d$ (the graded exterior differential) make $\cS^*[Q;Z]$
into a BGDA whose elements obey the relations
\be
\f\w\f' =(-1)^{|\f||\f'| +[\f][\f']}\f'\w \f, \qquad d(\f\w\f')=
d\f\w\f' +(-1)^{|\f|}\f\w d\f'.
\ee
Given the GDA $\cO^*Z$ of exterior forms on $Z$, there are the
monomorphism $\cO^*Z\to \cS^*[Q;Z]$ and the body epimorphism
$\cS^*[Q;Z]\to \cO^*Z$. The following facts are
essential.$^{9,11}$

\begin{lem} \label{v62} \mar{v62}
The BGDA $\cS^*[Q;Z]$ is a minimal differential calculus over
$\cA_Q$, i.e., it is generated by elements $df$, $f\in \cA_Q$.
\end{lem}

\begin{lem} \label{v30} \mar{v30}
Given a ring $R$, let $\cK$, $\cK'$ be $R$-rings and  $\cA$,
$\cA'$ the Grassmann algebras over  $\cK$ and $\cK'$,
respectively. Then a homomorphism (resp. a monomorphism) $\rho:
\cA\to \cA'$ yields a homomorphism (resp. a monomorphism) of the
minimal Chevalley--Eilenberg differential calculus over a $\Bbb
Z_2$-graded $R$-ring $\cA$ to that over $\cA'$ given by the map
$da \mapsto d(\rho(a))$, $a\in\cA$.
\end{lem}

One can think of elements of the BGDA $\cS^*[Q;Z]$ as being graded
exterior forms on $Z$ as follows. Given an open subset $U\subset
Z$, let $\cA_U$ be the Grassmann algebra of sections of the sheaf
$\gA_Q$ over $U$, and let $\cS^*[Q;U]$ be the Chevalley--Eilenberg
differential calculus over $\cA_U$. Given an open set $U'\subset
U$, the restriction morphisms $\cA_U\to\cA_{U'}$ yield a
homomorphism  of BGDAs $\cS^*[Q;U]\to \cS^*[Q;U']$. Thus,  we
obtain the presheaf $\{U,\cS^*[Q;U]\}$ of BGDAs on a manifold $Z$
and the sheaf $\gS^*[Q;Z]$ of BGDAs of germs of this presheaf.
Since $\{U,\cA_U\}$ is the canonical presheaf of $\gA_Q$, the
canonical presheaf of $\gS^*[Q;Z]$ is $\{U,\cS^*[Q;U]\}$. In
particular, $\cS^*[Q;Z]$ is the BGDA of global sections of the
sheaf $\gS^*[Q;Z]$, and  there is the restriction morphism
$\cS^*[Q;Z]\to \cS^*[Q;U]$ for any open $U\subset Z$. Due to this
morphism, elements of $\cS^*[Q;Z]$ can be written in the following
local form.

Given bundle coordinates $(z^A,q^a)$ on $Q$ and the corresponding
fiber basis $\{c^a\}$ for $Q^*\to X$, the tuple $(z^A, c^a)$ is
called a local basis for the graded manifold $(Z,\gA_Q)$.$^9$ With
respect to this basis, graded functions read
\mar{v23}\beq
f=\op\sum_{k=0} \frac1{k!}f_{a_1\ldots a_k}c^{a_1}\cdots c^{a_k},
\qquad f\in C^\infty(Z), \label{v23}
\eeq
where we omit the symbol of the exterior product of elements
$c^a$. Due to the canonical vertical splitting $VQ= Q\times Q$,
the fiber basis $\{\dr_a\}$ for the vertical tangent bundle $VQ\to
Q$ of $Q\to Z$ is the dual of $\{c^a\}$. Then graded derivations
take the local form $u= u^A\dr_A + u^a\dr_a$, where $u^A, u^a$ are
local graded functions. They act on graded functions (\ref{v23})
by the rule
\mar{cmp50'}\beq
u(f_{a\ldots b}c^a\cdots c^b)=u^A\dr_A(f_{a\ldots b})c^a\cdots c^b
+u^d f_{a\ldots b}\dr_d\rfloor (c^a\cdots c^b). \label{cmp50'}
\eeq
Relative to the dual local bases $\{dz^A\}$ for $T^*Z$ and
$\{dc^b\}$ for $Q^*$, graded one-forms read $\f=\f_A dz^A +
\f_adc^a$. The duality morphism is given by the interior product
\be
u\rfloor \f=u^A\f_A + (-1)^{[\f_a]}u^a\f_a, \qquad u\in \gd\cA_Q,
\qquad \f\in \cS^1[Q;Z].
\ee
The graded exterior differential reads
\be
d\f=dz^A\w \dr_A\f + dc^a\w \dr_a\f,
\ee
where the derivations $\dr_A$ and $\dr_a$ act on coefficients of
graded exterior forms by the formula (\ref{cmp50'}), and they are
graded commutative with the graded exterior forms $dz^A$ and
$dc^a$.

We define jets of odd fields as simple graded manifolds modelled
over jet bundles over $X$.$^{2,9}$ This definition differs from
the definition of jets of a graded commutative ring$^{11}$ and
that of jets of a graded fiber bundle,$^{14}$  but reproduces the
heuristic notion of jets of odd ghosts in Lagrangian BRST
theory.$^{7,15}$

Given a vector bundle $F\to X$, let us consider the simple graded
manifold $(J^rY,\gA_{F_r})$ whose body is $J^rY$ and the structure
bundle is the pull-back
\be
F_r=J^rY\op\times_XJ^rF
\ee
onto $J^rY$ of the jet bundle $J^rF\to X$, which is also a vector
bundle. Given the simple graded manifold
$(J^{r+1}Y,\gA_{F_{r+1}})$, there is an epimorphism of graded
manifolds
\be
(J^{r+1}Y,\gA_{F_{r+1}}) \to (J^rY,\gA_{F_r}).
\ee
It consists of the open surjection $\pi^{r+1}_r$ and the sheaf
monomorphism $\pi_r^{r+1*}\gA_{F_r}\to \gA_{F_{r+1}}$, where
$\pi_r^{r+1*}\gA_{F_r}$ is the pull-back onto $J^{r+1}Y$ of the
topological fiber bundle $\gA_{F_r}\to J^rY$. This sheaf
monomorphism induces the monomorphism of the canonical presheaves
$\ol \gA_{F_r}\to \ol \gA_{F_{r+1}}$, which associates to each
open subset $U\subset J^{r+1}Y$ the ring of sections of
$\gA_{F_r}$ over $\pi^{r+1}_r(U)$. Accordingly, there is the
monomorphism of $\Bbb Z_2$-graded rings $\cA_{F_r} \to
\cA_{F_{r+1}}$. By virtue of Lemmas \ref{v62} and \ref{v30}, this
monomorphism yields the monomorphism of BGDAs
\mar{v4}\beq
\cS^*[F_r;J^rY]\to \cS^*[F_{r+1};J^{r+1}Y]. \label{v4}
\eeq
As a consequence, we have the direct system of BGDAs
\mar{v6}\beq
\cS^*[Y\op\times_X F;Y]\ar \cS^*[F_1;J^1Y]\ar\cdots
\cS^*[F_r;J^rY]\ar\cdots, \label{v6}
\eeq
whose direct limit $\cS^*_\infty[F;Y]$  is a BGDA of all graded
differential forms $\f\in \cS^*[F_r;J^rY]$ on jet manifolds $J^rY$
modulo monomorphisms (\ref{v4}). The monomorphisms $\cO^*_rY\to
\cS^*[F_r;J^rY]$ provide the monomorphism $\cO^*_\infty Y\to
\cS^*_\infty[F;Y]$ of their direct limits. In particular,
$\cS^*_\infty[F;Y]$ is an $\cO^0_\infty Y$-algebra. Accordingly,
the body epimorphisms $\cS^*[F_r;J^rY]\to \cO^*_rY$ yield the
epimorphism of $\cO^0_\infty Y$-modules $\cS^*_\infty[F;Y]\to
\cO^*_\infty Y$.

If $Y\to X$ is an affine bundle, we recover the BGDA introduced in
Refs. \cite{jmp05,cmp04} by restricting the ring $\cO^0_\infty Y$
to its subring $\cP^0_\infty Y$ of polynomial functions, but now
elements of $\cS^*_\infty[F;Y]$ are graded exterior forms on
 $J^\infty Y$.
Indeed, let $\gS^*[F_r;J^rY]$ be the sheaf of BGDAs on $J^rY$ and
$\ol\gS^*[F_r;J^rY]$ its canonical presheaf whose elements are the
Chevalley--Eilenberg differential calculus over elements of the
presheaf $\ol\gA_{F_r}$. Then the presheaf monomorphisms $\ol
\gA_{F_r}\to \ol \gA_{F_{r+1}}$ yield the direct system of
presheaves
\mar{v15}\beq
\ol\gS^*[Y\times F;Y]\ar \ol\gS^*[F_1;J^1Y] \ar\cdots
\ol\gS^*[F_r;J^rY]  \ar\cdots, \label{v15}
\eeq
whose direct limit $\ol\gS_\infty^*[F;Y]$ is a presheaf of BGDAs
on the infinite order jet manifold $J^\infty Y$. Let
$\gQ^*_\infty[F;Y]$ be the sheaf of BGDAs of germs of the presheaf
$\ol\gS_\infty^*[F;Y]$.  The structure module
$\G(\gQ^*_\infty[F;Y])$ of sections of $\gQ^*_\infty[F;Y]$ is a
BGDA such that, given an element $\f\in \G(\gQ^*_\infty[F;Y])$ and
a point $z\in J^\infty Y$, there exist an open neighbourhood $U$
of $z$ and a graded exterior form $\f^{(k)}$ on some finite order
jet manifold $J^kY$ so that $\f|_U= \pi^{\infty*}_k\f^{(k)}|_U$.
In particular, there is the  monomorphism $\cS^*_\infty[F;Y]
\to\G(\gQ^*_\infty[F;Y])$.

Due to this monomorphism, one can restrict $\cS^*_\infty[F;Y]$ to
the coordinate chart (\ref{jet1}) and say that $\cS^*_\infty[F;Y]$
as an $\cO^0_\infty Y$-algebra is locally generated by  the
elements
\be
(1, c^a_\La,
dx^\la,\th^a_\La=dc^a_\La-c^a_{\la+\La}dx^\la,\th^i_\La=
dy^i_\La-y^i_{\la+\La}dx^\la), \qquad 0\leq |\La|.
\ee
 We agree to call $(y^i,c^a)$ the local basis for
$\cS^*_\infty[F;Y]$. Let the collective symbol $s^A$ stand for its
elements. Accordingly, the notation $s^A_\La$ and
$\th^A_\La=ds^A_\La- s^A_{\la+\La}dx^\la$ is introduced. For the
sake of simplicity, we further denote $[A]=[s^A]$.

The BGDA $\cS^*_\infty[F;Y]$ is decomposed into
$\cS^0_\infty[F;Y]$-modules $\cS^{k,r}_\infty[F;Y]$ of $k$-contact
and $r$-horizontal graded forms. Accordingly, the graded exterior
differential $d$ on $\cS^*_\infty[F;Y]$ falls into the sum
$d=d_H+d_V$ of the total and vertical differentials, where
\be
d_H(\f)=dx^\la\w d_\la(\f), \qquad d_\la = \dr_\la +
\op\sum_{0\leq|\La|} s^A_{\la+\La}\dr_A^\La.
\ee
Given the projector
\be
\vr=\op\sum_{k>0} \frac1k\ol\vr\circ h_k\circ h^n, \qquad
\ol\vr(\f)= \op\sum_{0\leq|\La|} (-1)^{\nm\La}\th^A\w
[d_\La(\dr^\La_A\rfloor\f)], \qquad \f\in \cS^{>0,n}_\infty[F;Y],
\ee
and the graded variational operator $\dl=\vr\circ d$, the BGDA
$\cS^*_\infty[F;Y]$ is split into the above mentioned
Grassmann-graded variational bicomplex.$^{7,8}$ We restrict our
consideration to its short variational subcomplex
\be
0\ar \Bbb R\ar \cS^0_\infty[F;Y]\ar^{d_H}\cS^{0,1}_\infty[F;Y]
\cdots \ar^{d_H} \cS^{0,n}_\infty[F;Y]\ar^\dl \bE_1, \quad
\bE_1=\vr(\cS^{1,n}_\infty[F;Y]).
\ee
One can think of its even elements
\mar{0709}\ben
&& L=\cL\om\in \cS^{0,n}_\infty[F;Y], \qquad \om=dx^1\w\cdots \w
dx^n,
\nonumber\\
&& \dl L= \th^A\w \cE_A\om=\op\sum_{0\leq|\La|}
 (-1)^{|\La|}\th^A\w d_\La (\dr^\La_A L)\om\in \bE_1 \label{0709}
\een
as being a graded Lagrangian and its Euler--Lagrange operator. A
pair $(\cS^*_\infty[F;Y], L)$ is further called a graded
Lagrangian system.

Let $\vt\in\gd \cS^0_\infty[F;Y]$ be a graded derivation of the
$\Bbb R$-ring $\cS^0_\infty[F;Y]$.$^{2,9}$ The interior product
$\vt\rfloor\f$  and the Lie derivative $\bL_\vt\f$,
$\f\in\cS^*_\infty[F;Y]$, are defined by the formulae
\be
&& \vt\rfloor \f=\vt^\la\f_\la + (-1)^{[\f_A]}\vt^A\f_A, \qquad
\f\in \cS^1_\infty[F;Y],\\
&& \vt\rfloor(\f\w\si)=(\vt\rfloor \f)\w\si
+(-1)^{|\f|+[\f][\vt]}\f\w(\vt\rfloor\si), \qquad \f,\si\in
\cS^*_\infty[F;Y], \\
&& \bL_\vt\f=\vt\rfloor d\f+ d(\vt\rfloor\f), \qquad
\bL_\vt(\f\w\si)=\bL_\vt(\f)\w\si
+(-1)^{[\vt][\f]}\f\w\bL_\vt(\si).
\ee
A graded derivation $\vt$
 is said to be contact if the Lie
derivative $\bL_\vt$ preserves the ideal of contact graded forms
of the BGDA $\cS^*_\infty[F;Y]$. With respect to the local basis
$\{s^A\}$ for the BGDA $\cS^*_\infty[F;Y]$, any contact graded
derivation takes the form
\be
\vt=\vt_H+\vt_V=\vt^\la d_\la + (\vt^A\dr_A +\op\sum_{0<|\La|}
d_\La\vt^A\dr_A^\La),
\ee
where the tuple of graded derivations $\{\dr_\la,\dr^\La_A\}$ is
 the dual of the tuple $\{dx^\la, ds^A_\La\}$ of
generating elements of the $\cS^0_\infty[F;Y]$-algebra
$\cS^*_\infty[F;Y]$, and $\vt^\la$, $\vt^A$ are local graded
functions.

We restrict our consideration to vertical contact graded
derivations
\mar{0672}\beq
\vt=\op\sum_{0\leq|\La|} d_\La\up^A\dr_A^\La. \label{0672}
\eeq
Such a derivation is completely determined by its first summand
\mar{0673}\beq
\up=\up^A(x^\la,s^A_\La)\dr_A, \qquad 0\leq|\La|\leq k,
\label{0673}
\eeq
called a generalized graded vector field. It is said to be
nilpotent if
\be
\bL_\vt(\bL_\vt\f)= \op\sum_{0\leq|\Si|,0\leq|\La|}
(\up^B_\Si\dr^\Si_B(\up^A_\La)\dr^\La_A +
(-1)^{[B][\up^A]}\up^B_\Si\up^A_\La\dr^\Si_B \dr^\La_A)\f=0
\ee
for any horizontal graded form $\f\in \cS^{0,*}_\infty[F;Y]$. One
can show that $\vt$ (\ref{0672}) is nilpotent only if it is odd
and iff all $\up^A$ obey the equality
\mar{0688}\beq
\vt(\up^A)=\op\sum_{0\leq|\Si|} \up^B_\Si\dr^\Si_B(\up^A)=0.
\label{0688}
\eeq

For the sake of simplicity, the common symbol further stands for a
generalized vector field (\ref{0673}), the contact graded
derivation (\ref{0672}) determined by this field and the Lie
derivative $\bL_\vt$. We agree to call all these operators the
graded derivation of the BGDA $\cS^*_\infty[F;Y]$.

\bigskip
\bigskip

\noindent {\bf III. NOETHER IDENTITIES IN A GENERAL SETTING}
\bigskip

Given a graded Lagrangian system $(\cS^*_\infty[F;Y],L)$, let us
construct the manifested Koszul--Tate complex of its Noether
identities.

The main ingredient in this construction is BGDAs of the following
type. Given a vector bundle $E\to X$, let us consider the BGDA
$\cS^*_\infty[F;E_Y]$, where $E_Y$ denotes the pull-back of $E$
onto $Y$. There are monomorphisms of $\cO^0_\infty Y$-algebras
\be
\cS^*_\infty[F;Y]\to \cS^*_\infty[F;E_Y], \qquad \cO^*_\infty E\to
\cS^*_\infty[F;E_Y],
\ee
whose images contain the common subalgebra $\cO^*_\infty Y$. Let
us consider: (i) the subring $\cP^0_\infty E_Y\subset \cO^0_\infty
E_Y$ of polynomial functions in fiber coordinates  of the vector
bundles $J^rE_Y\to J^rY$, $r\in \Bbb N$, (ii) the corresponding
subring $\cP^0_\infty[F;E_Y]\subset\cS^0_\infty[F;E_Y]$ of graded
functions with polynomial coefficients belonging to $\cP^0_\infty
E_Y$, (iii) the  subalgebra  $\cP^*_\infty[F;Y;E]$ of the BGDA
$\cS^*_\infty[F;E_Y]$ over the subring $\cP^0_\infty[F;E_Y]$.
Given vector bundles $V,V',E,E'$ over $X$, we further use the
notation
\be
\cP^*_\infty[V'V;F;Y;EE']= \cP^*_\infty[V'\op\times_X
V\op\times_XF;Y;E\op\times_X E'].
\ee
By a density-dual of a vector bundle $E\to X$ is meant
\be
\ol E^*=E^*\op\ot_X\op\w^n T^*X.
\ee

For the sake of simplicity, we restrict our consideration to
Lagrangian systems where a fiber bundle $Y\to X$ of even fields
admits the vertical splitting $VY=Y\times W$, where $W$ is a
vector bundle over $X$. This is case of almost all field models.
In a general setting, one must require that transition functions
of fiber bundles over $Y$ under consideration do not vanish
on-shell. Let $\ol Y^*$ denote the density-dual of $W$ in the
above mentioned vertical splitting.

\begin{prop} \label{v120} \mar{v120} One can associate to a graded
Lagrangian system $(\cS^*_\infty[F;Y],L)$, a chain complex whose
boundaries vanish on shell (see the complex (\ref{v42}) below).
\end{prop}

\begin{proof}
  Let us extend the BGDA $\cS^*_\infty[F;Y]$ to
the BGDA $\cP^*_\infty[\ol Y^*;F;Y;\ol F^*]$ whose local basis is
$\{s^A, \ol s_A\}$, where  $[\ol s_A]=([A]+1){\rm mod}\,2$.
Following the terminology of Lagrangian BRST theory,$^{2,5}$ we
call $\ol s_A$ the antifields of antifield number 1. The BGDA
$\cP^0_\infty[\ol Y^*;F;Y;\ol F^*]$ is provided with the nilpotent
graded derivation $\ol\dl=\rdr^A \cE_A$, where $\cE_A$ are the
graded variational derivatives (\ref{0709}) and the tuple of
graded right derivations $\{\rdr^{\La A}\}$ is the dual of the
tuple of contact graded forms $\{\th_{\La A}\}$. Because of the
expression (\ref{0709}) for $\dl L$, it is convenient to deal with
a graded derivation $\ol\dl$ acting on graded functions and forms
$\f$ on the right by the rule
\be
\ol\dl(\f)=d\f\lfloor\ol\dl +d(\f\lfloor\ol\dl), \qquad
\ol\dl(\f\w\f')=(-1)^{[\f']}\ol\dl(\f)\w\f'+ \f\w\ol\dl(\f').
\ee
We call $\ol\dl$ the Koszul--Tate differential. Let us consider
the module $\cP^{0,n}_\infty[\ol Y^*;F;Y;\ol F^*]$ of graded
densities. It is split into the chain complex
\mar{v42}\beq
0\lto \cS^{0,n}_\infty[F;Y] \llr^{\ol\dl} \cP^{0,n}_\infty[\ol
Y^*;F;Y;\ol F^*]_1\cdots \llr^{\ol\dl} \cP^{0,n}_\infty[\ol
Y^*;F;Y;\ol F^*]_k \cdots \label{v42}
\eeq
graded by the antifield number of its elements. It is readily
observed that the boundaries of the complex (\ref{v42}) vanish
on-shell.
\end{proof}

Note that the homology groups $H_*(\ol\dl)$ of the complex
(\ref{v42}) are $\cS^0_\infty[F;Y]$-modules, but these modules
fail to be torsion-free. Indeed, given a cycle $\f\in
\cP^{0,n}_\infty[\ol Y^*;F;Y;\ol F^*]_k$ and an element
$f=\ol\dl\si$ of the ring $\cS^0_\infty[F;Y]\subset
\cP^0_\infty[\ol Y^*;F;Y;\ol F^*]$, we obtain that
$f\f=\ol\dl(\si\f)$ is a boundary. Therefore, one can not apply
the K\"unneth formula to the homology of this complex, though any
its term $\cP^{0,n}_\infty[\ol Y^*;F;Y;\ol F^*]_k$ is isomorphic
to the graded commutative $k$-tensor product of the
$\cS^0_\infty[F;Y]$-module $\cP^{0,n}_\infty[\ol Y^*;F;Y;\ol
F^*]_1$.

The homology $H_0(\ol\dl)$ of the complex (\ref{v42}) is not
trivial, but this homology and the higher ones $H_{k\geq
2}(\ol\dl)$ are not essential for our consideration. Therefore, we
replace the complex (\ref{v42}) with the finite one
\mar{v042}\beq
0\lto \im\ol\dl \llr^{\ol\dl} \cP^{0,n}_\infty[\ol Y^*;F;Y;\ol
F^*]_1 \llr^{\ol\dl} \cP^{0,n}_\infty[\ol Y^*;F;Y;\ol F^*]_2
\label{v042}
\eeq
of graded densities of antifield number $k\leq 2$. It is exact at
$\im\ol\dl$, and  its first homology coincides with that of  the
complex (\ref{v42}). Let us consider this homology.

A generic one-chain of the complex (\ref{v042}) takes the form
\mar{0712}\beq
\Phi= \op\sum_{0\leq|\La|} \Phi^{A,\La}\ol s_{\La A} \om, \qquad
\Phi^{A,\La}\in \cS^0_\infty[F;Y], \label{0712}
\eeq
and the cycle condition $\ol\dl \Phi=0$ reads
\mar{0713}\beq
\op\sum_{0\leq|\La|} \Phi^{A,\La} d_\La \cE_A \om=0. \label{0713}
\eeq
 One can think of this equality as being a reduction condition on the
graded variational derivatives $\cE_A$. Conversely, any reduction
condition of form (\ref{0713}) comes from some cycle (\ref{0712}).
The reduction condition (\ref{0713}) is trivial  if a cycle is a
boundary, i.e., it takes the form
\mar{v44}\beq
\Phi= \op\sum_{0\leq|\La|,|\Si|} T^{(A\La)(B\Si)}d_\Si\cE_B\ol
s_{\La A}\om, \qquad T^{(A\La)(B\Si)}=-(-1)^{[A][B]}
T^{(B\Si)(A\La)}. \label{v44}
\eeq

\begin{defi} \label{v121} \mar{v121}
A graded Lagrangian system is called degenerate if there exist
non-trivial reduction conditions
 (\ref{0713}), called Noether
identities.
\end{defi}

One can say something more if the $\cS^0_\infty[F;Y]$-module
$H_1(\ol \dl)$ is finitely generated, i.e., it possesses the
following particular structure. There are elements $\Delta\in
H_1(\ol \dl)$ making up a $\Bbb Z_2$-graded projective
$C^\infty(X)$-module $\cC_{(0)}$ of finite rank which, by virtue
of the Serre--Swan theorem, is isomorphic to the module of
sections of the product $\ol V^*\op\times_X \ol E^*$ of the
density-duals of some vector bundles $V\to X$ and $E\to X$. Let
$\{\Delta_r\}$ be local bases for this $C^\infty(X)$-module. Every
element $\Phi\in H_1(\ol \dl)$ factorizes
\mar{v63,71}\ben
&& \Phi= \op\sum_{0\leq|\Xi|} G^{r,\Xi} d_\Xi \Delta_r\om, \qquad
G^{r,\Xi}\in
\cS^0_\infty[F;Y], \label{v63}\\
&&\Delta_r=\op\sum_{0\leq|\La|} \Delta_r^{A,\La}\ol s_{\La A},
\qquad \Delta_r^{A,\La}\in \cS^0_\infty[F;Y], \label{v71}
\een
via elements of $\cC_{(0)}$, i.e., any Noether identity
(\ref{0713}) is a corollary of Noether identities
\mar{v64}\beq
 \op\sum_{0\leq|\La|} \Delta_r^{A,\La} d_\La \cE_A=0.
\label{v64}
\eeq
Clearly, the factorization (\ref{v63}) is independent of
specification of local bases $\{\Delta_r\}$. We say that the
Noether identities  (\ref{v64}) are complete, and call $\Delta\in
\cC_{(0)}$ the Noether operators. Note that, being representatives
of $H_1(\ol \dl)$, the graded densities $\Delta_r$ (\ref{v71}) are
not $\ol\dl$-exact.

\begin{prop} \label{v137} \mar{v137}
If the homology $H_1(\ol\dl)$ of the complex (\ref{v042}) is
finitely generated, this complex can be extended to a one-exact
complex  with a boundary operator whose nilpotency conditions are
just complete Noether identities (see the complex (\ref{v66})
below).
\end{prop}

\begin{proof}
Let us extend the BGDA $\cP^*_\infty[\ol Y^*;F;Y;\ol F^*]$ to  the
BGDA $\cP^*_\infty[\ol E^*\ol Y^*;F;Y;\ol F^*\ol V^*]$ possessing
the local basis $\{s^A,\ol s_A, \ol c_r\}$, where $[\ol
c_r]=([\Delta_r]+1){\rm mod}\,2$ and Ant$[\ol c]=2$. It is
provided with the nilpotent graded derivation
\mar{v204}\beq
\dl_0=\ol\dl + \rdr^ r\Delta_r, \label{v204}
\eeq
called the extended Koszul--Tate differential. Its nilpotency
conditions (\ref{0688}) are equivalent to the complete Noether
identities (\ref{v64}). Then the module $\cP^{0,n}_\infty[\ol
E^*\ol Y^*;F;Y;\ol F^*\ol V^*]_{\leq 3}$ of graded densities of
antifield number Ant$[\f]\leq 3$ is split into the chain complex
\mar{v66}\ben
&&0\lto \im\ol\dl \llr^{\ol\dl} \cP^{0,n}_\infty[\ol Y^*;F;Y;\ol
F^*]_1\llr^{\dl_0}
\cP^{0,n}_\infty[\ol E^*\ol Y^*;F;Y;\ol F^*\ol V^*]_2 \label{v66}\\
&& \qquad \llr^{\dl_0} \cP^{0,n}_\infty[\ol E^*\ol Y^*;F;Y;\ol
F^*\ol V^*]_3. \nonumber
\een
Let $H_*(\dl_0)$ denote its homology. We have
$H_0(\dl_0)=H_0(\ol\dl)=0$. Furthermore, any one-cycle $\Phi$ up
to a boundary takes the form (\ref{v63}) and, therefore, it is a
$\dl_0$-boundary
\be
\Phi= \op\sum_{0\leq|\Si|} G^{r,\Xi} d_\Xi \Delta_r\om
=\dl_0(\op\sum_{0\leq|\Si|} G^{r,\Xi}\ol c_{\Xi r}\om).
\ee
Hence, $H_1(\dl_0)=0$, i.e., the complex (\ref{v66}) is one-exact.
\end{proof}

\bigskip
\bigskip

\noindent {\bf IV. THE KOSZUL-TATE COMPLEX OF NOETHER IDENTITIES}
\bigskip

Turn now to the homology $H_2(\dl_0)$ of the complex (\ref{v66}).
A generic two-chain  reads
\mar{v77}\ben
&& \Phi= G + H= \op\sum_{0\leq|\La|} G^{r,\La}\ol c_{\La r}\om +
\op\sum_{0\leq|\La|,|\Si|} H^{(A,\La)(B,\Si)}\ol s_{\La A}\ol
s_{\Si B}\om,
\label{v77}\\
&& G^{r,\La} \in \cS^0_\infty[F;Y], \qquad H^{(A,\La)(B,\Si)}\cV
\in \cS^0_\infty[F;Y], \qquad \cV\in\cO^nX. \nonumber
\een
The cycle condition $\dl_0 \Phi=0$ takes the form
\mar{v79}\beq
 \op\sum_{0\leq|\La|} G^{r,\La}d_\La\Delta_r\om +\ol\dl H=0.
\label{v79}
\eeq
One can think of this equality as being the reduction condition on
the Noether operators (\ref{v71}). Conversely, let
\be
\Phi=\op\sum_{0\leq|\La|} G^{r,\La}\ol c_{\La r}\om\in
\cP^{0,n}_\infty[\ol E^*\ol Y^*;F;Y;\ol F^*\ol V^*]_2
\ee
be a graded density such that the reduction condition (\ref{v79})
holds. Obviously, it is a cycle condition of the two-chain
(\ref{v77}). The reduction condition (\ref{v79}) is trivial either
if a two-cycle $\Phi$ (\ref{v77}) is a boundary or its summand $G$
vanishes on-shell.

\begin{defi} \label{v133} \mar{v133}
A degenerate graded Lagrangian system in Proposition \ref{v137} is
said to be one-stage reducible if there exist non-trivial
reduction conditions (\ref{v79}), called the first-stage Noether
identities.
\end{defi}

\begin{prop} \label{v134} \mar{v134}
First-stage Noether identities can be identified to nontrivial
elements of the homology $H_2(\dl_0)$ iff any $\ol\dl$-cycle
$\f\in \cP^{0,n}_\infty[\ol Y^*;F;Y;\ol F^*]_2$ is a
$\dl_0$-boundary.
\end{prop}

\begin{proof}
It suffices to show that, if the summand $G$ of a two-cycle $\Phi$
(\ref{v77}) is $\ol\dl$-exact, then $\Phi$ is a boundary. If
$G=\ol\dl \Psi$, then
\mar{v169}\beq
\Phi=\dl_0\Psi +(\ol \dl-\dl_0)\Psi + H. \label{v169}
\eeq
The cycle condition reads
\be
\dl_0\Phi=\ol\dl((\ol\dl-\dl_0)\Psi + H)=0.
\ee
Then $(\ol \dl-\dl_0)\Psi + H$ is $\dl_0$-exact since any
$\ol\dl$-cycle $\f\in \cP^{0,n}_\infty[\ol Y^*;F;Y;\ol F^*]_2$, by
assumption, is a $\dl_0$-boundary. Consequently, $\Phi$
(\ref{v169}) is $\dl_0$-exact. Conversely, let $\Phi\in
\cP^{0,n}_\infty[\ol Y^*;F;Y;\ol F^*]_2$ be an arbitrary
$\ol\dl$-cycle.
 The cycle condition reads
\mar{v100}\beq
\ol\dl\Phi= 2\Phi^{(A,\La)(B,\Sigma)}\ol s_{\La A} \ol\dl\ol
s_{\Sigma B}\om= 2\Phi^{(A,\La)(B,\Sigma)}\ol s_{\La A} d_\Si
\cE_B\om=0. \label{v100}
\eeq
It follows that $\Phi^{(A,\La)(B,\Sigma)} \ol\dl\ol s_{\Sigma
B}=0$ for all indices $(A,\La)$. Omitting a $\ol\dl$-boundary
term, we obtain
\be
\Phi^{(A,\La)(B,\Sigma)} \ol s_{\Sigma B}= G^{(A,\La)(r,\Xi)}d_\Xi
\Delta_r.
\ee
Hence, $\Phi$ takes the form
\mar{v135}\beq
\Phi=G'^{(A,\La)(r,\Xi)} d_\Xi\Delta_r \ol s_{\La A}\om.
\label{v135}
\eeq
We can associate to it the three-chain
\be
\Psi= G'^{(A,\La)(r,\Xi)} \ol c_{\Xi r} \ol s_{\La A}\om
\ee
such that
\be
\dl_0\Psi=\Phi +\si = \Phi + G''^{(A,\La)(r,\Xi)}d_\La\cE_A \ol
c_{\Xi r} \om.
\ee
Owing to the equality $\ol\dl\Phi=0$, we have $\dl_0\si=0$. Since
$\si$ is $\ol\dl$-exact, it by assumption is $\dl_0$-exact, i.e.,
$\si=\dl_0\psi$.  Then we obtain that $\Phi=\dl_0\Psi -\dl_0\psi$.
\end{proof}

\begin{lem} \label{v200} \mar{v200}
It is easily justified that a two-cycle $\Phi\in
\cP^{0,n}_\infty[\ol Y^*;F;Y;\ol F^*]_2$ is  $\dl_0$-exact iff
$\Phi$ up to a $\ol\dl$-boundary takes the form
\mar{v140}\beq
\Phi= \op\sum_{0\leq |\La|, |\Si|} G'^{(r,\Si)(r',\La)}
d_\Si\Delta_r d_\La\Delta_{r'}\om. \label{v140}
\eeq
\end{lem}

If the condition of Proposition \ref{v134} (called the
two-homology regularity condition) is satisfied, let us assume
that the first-stage Noether identities are finitely generated as
follows. There are elements $\Delta_{(1)}\in H_2(\dl_0)$ making up
a $\Bbb Z_2$-graded projective $C^\infty(X)$-module $\cC_{(1)}$ of
finite rank which is isomorphic to the module of sections of the
product $\ol V^*_1\op\times_X \ol E^*_1$  of the density-duals of
some vector bundles $V_1\to X$ and $E_1\to X$. Let
$\{\Delta_{r_1}\}$ be local bases for this $C^\infty(X)$-module.
Every element $\Phi\in H_2(\dl_0)$ factorizes
\mar{v80,1}\ben
&& \Phi= \op\sum_{0\leq|\Xi|} \Phi^{r_1,\Xi} d_\Xi
\Delta_{r_1}\om, \qquad \Phi^{r_1,\Xi}\in
\cS^0_\infty[F;Y], \label{v80}\\
&&\Delta_{r_1}=G_{r_1}+ h_{r_1}=\op\sum_{0\leq|\La|}
\Delta_{r_1}^{r,\La}\ol c_{\La r} + h_{r_1}, \qquad
 h_{r_1}\om\in
\cP^{0,n}_\infty[\ol Y^*;F;Y;\ol F^*], \label{v81}
\een
via elements of $\cC_{(1)}$, i.e., any first-stage Noether
identity (\ref{v79}) results from the equalities
\mar{v82}\beq
 \op\sum_{0\leq|\La|} \Delta_{r_1}^{r,\La} d_\La \Delta_r +\ol\dl
h_{r_1} =0, \label{v82}
\eeq
called the complete first-stage Noether identities. Elements of
$\cC_{(1)}$ are called the first-stage Noether operators. Note
that first summands $G_{r_1}$ of  operators $\Delta_{r_1}$
(\ref{v81}) are not $\ol\dl$-exact.

\begin{prop} \label{v139} \mar{v139} Given a reducible degenerate
Lagrangian system, let the associated one-exact complex
(\ref{v66}) obey the two-homology regularity condition and let its
homology $H_2(\dl_0)$ (first-stage Noether identities) be finitely
generated. Then this complex is extended to the two-exact one
with a boundary operator whose nilpotency conditions are
equivalent to complete Noether and first-stage Noether identities
(see the complex (\ref{v87}) below).
\end{prop}

\begin{proof}
Let us consider the BGDA $\cP^*_\infty[\ol E^*_1\ol E^*\ol
Y^*;F;Y;\ol F^*\ol V^*\ol V^*_1]$ with the local basis $\{s^A,\ol
s_A, \ol c_r, \ol c_{r_1}\}$, where $[\ol
c_{r_1}]=([\Delta_{r_1}]+1){\rm mod}\,2$ and Ant$[\ol c_{r_1}]=3$.
It can be provided the first-stage Koszul--Tate differential
defined as the nilpotent graded derivation
\mar{v205}\beq
\dl_1=\dl_0 + \rdr^{r_1} \Delta_{r_1}. \label{v205}
\eeq
Its nilpotency conditions (\ref{0688}) are equivalent to complete
Noether identities (\ref{v64}) and complete first-stage Noether
identities (\ref{v82}). Then the module $\cP^{0,n}_\infty[\ol
E^*_1\ol E^*\ol Y^*;F;Y;\ol F^*\ol V^*\ol V^*_1]_{\leq 4}$ of
graded densities of antifield number Ant$[\f]\leq 4$ is split into
the chain complex
\mar{v87}\ben
&&0\lto \im\ol\dl \llr^{\ol\dl} \cP^{0,n}_\infty[\ol Y^*;F;Y;\ol
F^*]_1\llr^{\dl_0} \cP^{0,n}_\infty[\ol E^*\ol Y^*;F;Y;\ol F^*\ol
V^*]_2\llr^{\dl_1}
\label{v87}\\
&& \qquad \cP^{0,n}_\infty[\ol E^*_1\ol E^*\ol Y^*;F;Y;\ol F^*\ol
V^*\ol V^*_1]_3
 \llr^{\dl_1}
\cP^{0,n}_\infty[\ol E^*_1\ol E^*\ol Y^*;F;Y;\ol F^*\ol V^*\ol
V^*_1]_4. \nonumber
\een
Let $H_*(\dl_1)$ denote its homology. It is readily observed that
\be
H_0(\dl_1)=H_0(\ol\dl), \qquad H_1(\dl_1)=H_1(\dl_0)=0.
\ee
By virtue of the expression (\ref{v80}), any two-cycle of the
complex (\ref{v87}) is a boundary
\be
 \Phi= \op\sum_{0\leq|\Xi|} \Phi^{r_1,\Xi} d_\Xi \Delta_{r_1}\om
=\dl_1(\op\sum_{0\leq|\Xi|} \Phi^{r_1,\Xi} \ol c_{\Xi r_1})\om.
\ee
It follows that $H_2(\dl_1)=0$, i.e., the complex (\ref{v87}) is
two-exact.
\end{proof}

If the third homology $H_3(\dl_1)$ of the complex (\ref{v87}) is
not trivial, there are reduction conditions on the first-stage
Noether operators, and so on. Iterating the arguments, we come to
the following.

Let $(\cS^*_\infty[F;Y],L)$ be a degenerate graded Lagrangian
system whose Noether identities are finitely generated. In
accordance with Proposition \ref{v137}, we associates to it the
one-exact chain complex (\ref{v66}). Given an integer $N\geq 1$,
let $V_1,\ldots, V_N, E_1, \ldots, E_N$ be some vector bundles
over $X$ and
\mar{v91}\beq
\cP^*_\infty\{N\}=\cP^*_\infty[\ol E^*_N\cdots\ol E^*_1\ol E^*\ol
Y^*;F;Y;\ol F^*\ol V^*\ol V^*_1\cdots\ol V_N^*] \label{v91}
\eeq
a BGDA with local bases $\{s^A,\ol s_A, \ol c_r, \ol c_{r_1},
\ldots, \ol c_{r_N}\}$ graded by antifield numbers Ant$[\ol
c_{r_k}]=k+2$. Let $k=-1,0$ further stand for $\ol s_A$ and $\ol
c_r$, respectively. We assume that:

(i) the BGDA $\cP^*_\infty\{N\}$ (\ref{v91}) is provided with a
nilpotent graded derivation
\mar{v92,'}\ben
&&\dl_N=\dl_0 + \op\sum_{1\leq k\leq N}\rdr^{r_k} \Delta_{r_k},
\label{v92}\\
&& \Delta_{r_k}=G_{r_k} + h_{r_k}= \op\sum_{0\leq|\La|}
\Delta_{r_k}^{r_{k-1},\La}\ol c_{\La r_{k-1}} + \op\sum_{0\leq
\Si, 0\leq\Xi}(h_{r_k}^{(A,\Xi)(r_{k-2},\Si)}\ol s_{\Xi A}\ol
c_{\Si r_{k-2}}+...), \label{v92'}
\een
of antifield number -1;

(ii) the module $\cP^{0,n}_\infty\{N\}_{\leq N+3}$ of graded
densities of antifield number Ant$[\f]\leq N+3$ is split into the
$(N+1)$-exact chain complex
\mar{v94}\ben
&&0\lto \im \ol\dl \llr^{\ol\dl} \cP^{0,n}_\infty[\ol Y^*;F;Y;\ol
F^*]_1\llr^{\dl_0} \cP^{0,n}_\infty\{0\}_2\llr^{\dl_1}
\cP^{0,n}_\infty\{1\}_3\cdots
\label{v94}\\
&& \qquad
 \llr^{\dl_{N-1}} \cP^{0,n}_\infty\{N-1\}_{N+1}
\llr^{\dl_N} \cP^{0,n}_\infty\{N\}_{N+2}\llr^{\dl_N}
\cP^{0,n}_\infty\{N\}_{N+3}, \nonumber
\een
which satisfies the $(N+1)$-homology regularity condition in
accordance with forthcoming Definition \ref{v155}.

\begin{defi} \label{v155} \mar{v155} One says that the complex (\ref{v94})
obeys the $(N+1)$-homology regularity condition if any
$\dl_{k<N-1}$-cycle $\f\in \cP_\infty^{0,n}\{k\}_{k+3}\subset
\cP_\infty^{0,n}\{k+1\}_{k+3}$ is a $\dl_{k+1}$-boundary.
\end{defi}

\begin{rem}
The $(N+1)$-exactness of the complex (\ref{v94}) implies that any
$\dl_{k<N-1}$-cycle $\f\in \cP_\infty^{0,n}\{k\}_{k+3}$, $k<N$, is
a $\dl_{k+2}$-boundary, but not necessary a $\dl_{k+1}$-one.
\end{rem}

If $N=1$, the complex $\cP^{0,n}_\infty\{1\}_{\leq 4}$ (\ref{v94})
restarts the complex (\ref{v87}) associated to a first-stage
reducible graded Lagrangian system by virtue of Proposition
\ref{v139}. Therefore, we agree to call $\dl_N$ (\ref{v92}) the
$N$-stage Koszul--Tate differential. Its nilpotency implies
complete Noether identities (\ref{v64}), first-stage Noether
identities (\ref{v82}) and the equalities
\mar{v93}\beq
\op\sum_{0\leq|\La|} \Delta_{r_k}^{r_{k-1},\La}d_\La
(\op\sum_{0\leq|\Si|} \Delta_{r_{k-1}}^{r_{k-2},\Si}\ol c_{\Si
r_{k-2}}) + \ol\dl(\op\sum_{0\leq \Si,
0\leq\Xi}h_{r_k}^{(A,\Xi)(r_{k-2},\Si)}\ol s_{\Xi A}\ol c_{\Si
r_{k-2}})=0,  \label{v93}
\eeq
for $k=2,\ldots,N$. One can think of the equalities (\ref{v93}) as
being complete $k$-stage Noether identities because of their
properties which we will justify in the case of $k=N+1$.
Accordingly, $\Delta_{r_k}$ (\ref{v92'}) are said to be the
$k$-stage Noether operators.

Let us consider the $(N+2)$-homology of the complex (\ref{v94}). A
generic $(N+2)$-chain $\Phi\in \cP^{0,n}_\infty\{N\}_{N+2}$ takes
the form
\mar{v156}\beq
\Phi= G + H= \op\sum_{0\leq|\La|} G^{r_N,\La}\ol c_{\La r_N}\om +
\op\sum_{0\leq \Si, 0\leq\Xi}(H^{(A,\Xi)(r_{N-1},\Si)}\ol s_{\Xi
A}\ol c_{\Si r_{N-1}}+...)\om. \label{v156}
\eeq
Let it be a cycle. The cycle condition $\dl_N\Phi=0$ implies the
equality
\mar{v145}\beq
\op\sum_{0\leq|\La|} G^{r_N,\La}d_\La (\op\sum_{0\leq|\Si|}
\Delta_{r_N}^{r_{N-1},\Si}\ol c_{\Si r_{N-1}}) +
\ol\dl(\op\sum_{0\leq \Si, 0\leq\Xi}H^{(A,\Xi)(r_{N-1},\Si)}\ol
s_{\Xi A}\ol c_{\Si r_{N-1}})=0.  \label{v145}
\eeq
One can think of this equality as being the reduction condition on
the $N$-stage Noether operators (\ref{v92'}). Conversely, let
\be
\Phi= \op\sum_{0\leq|\La|} G^{r_N,\La}\ol c_{\La r_N}\om \in
\cP^{0,n}_\infty\{N\}_{N+2}
\ee
be a graded density such that the reduction condition (\ref{v145})
holds. Then this reduction condition can be extended to a cycle
one as follows. It is brought into the form
\be
\dl_N(\op\sum_{0\leq|\La|} && G^{r_N,\La}\ol c_{\La r_N} +
\op\sum_{0\leq \Si, 0\leq\Xi}H^{(A,\Xi)(r_{N-1},\Si)}\ol
s_{\Xi A}\ol c_{\Si r_{N-1}})=\\
&& \qquad  -\op\sum_{0\leq|\La|} G^{r_N,\La}d_\La h_{r_N}
+\op\sum_{0\leq \Si, 0\leq\Xi}H^{(A,\Xi)(r_{N-1},\Si)}\ol s_{\Xi
A}d_\Si \Delta_{r_{N-1}}.
\ee
A glance at the expression (\ref{v92'}) shows that the term in the
right-hand side of this equality belongs to
$\cP^{0,n}_\infty\{N-2\}_{N+1}$. It is a $\dl_{N-2}$-cycle and,
consequently, a $\dl_{N-1}$-boundary $\dl_{N-1}\Psi$ in accordance
with the $(N+1)$-homology regularity condition. Then the reduction
condition (\ref{v145}) is a $\ol c_{\Si r_{N-1}}$-dependent part
of the cycle condition
\be
\dl_N(\op\sum_{0\leq|\La|} && G^{r_N,\La}\ol c_{\La r_N} +
\op\sum_{0\leq \Si, 0\leq\Xi}H^{(A,\Xi)(r_{N-1},\Si)}\ol s_{\Xi
A}\ol c_{\Si r_{N-1}} -\Psi)=0,
\ee
but $\dl_N\Psi$ does not make a contribution to this reduction
condition.

Being a cycle condition, the reduction condition (\ref{v145}) is
trivial either if a cycle $\Phi$ (\ref{v156}) is a
$\dl_N$-boundary or its summand $G$ is $\ol\dl$-exact, i.e., it is
a boundary, too, as we have stated above. Then Definition
\ref{v133} can be generalized as follows.

\begin{defi} \label{v202} \mar{v202}
A degenerate graded Lagrangian system is said to be $(N+1)$-stage
reducible if there exist non-trivial reduction conditions
(\ref{v145}), called the $(N+1)$-stage Noether identities.
\end{defi}

\begin{theo} \label{v163} \mar{v163}
(i) The $(N+1)$-stage Noether identities can be identified to
nontrivial elements of the homology $H_{N+2}(\dl_N)$ of the
complex (\ref{v94}) iff this homology obeys the $(N+2)$-homology
regularity condition. (ii) If the homology $H_{N+2}(\dl_N)$ is
finitely generated as defined below, the complex (\ref{v94})
admits an $(N+2)$-exact extension.
\end{theo}

\begin{proof}
(i) The $(N+2)$-homology regularity condition implies that any
$\dl_{N-1}$-cycle $\Phi\in \cP_\infty^{0,n}\{N-1\}_{N+2}\subset
\cP_\infty^{0,n}\{N\}_{N+2}$ is a $\dl_N$-boundary. Therefore, if
$\Phi$ (\ref{v156}) is a representative of a nontrivial element of
$H_{N+2}(\dl_N)$, its summand $G$ linear in $\ol c_{\La r_N}$ does
not vanish. Moreover, it is not a $\ol\dl$-boundary. Indeed, if
$\Phi=\ol\dl \Psi$, then
\mar{v172}\beq
\Phi=\dl_N\Psi +(\ol \dl-\dl_N)\Psi + H. \label{v172}
\eeq
The cycle condition takes the form
\be
\dl_N\Phi=\dl_{N-1}((\ol\dl-\dl_N)\Psi + H)=0.
\ee
Hence, $(\ol \dl-\dl_N)\Psi + H$ is $\dl_N$-exact since any
$\dl_{N-1}$-cycle $\f\in \cP_\infty^{0,n}\{N-1\}_{N+2}$ is a
$\dl_N$-boundary. Consequently, $\Phi$ (\ref{v172}) is a boundary.
If the $(N+2)$-homology regularity condition does not hold,
trivial reduction conditions (\ref{v145}) also come from
nontrivial elements of the homology $H_{N+2}(\dl_N)$. (ii) Let the
$(N+1)$-stage Noether identities be finitely generated. Namely,
there exist elements $\Delta_{(N+1)}\in H_{N+2}(\dl_N)$ making up
a $\Bbb Z_2$-graded projective $C^\infty(X)$-module $\cC_{(N+1)}$
of finite rank which is isomorphic to the module of sections of
the product $\ol V^*_{N+1}\op\times_X \ol E^*_{N+1}$ of the
density-duals of some vector bundles $V_{N+1}\to X$ and
$E_{N+1}\to X$. Let $\{\Delta_{r_{N+1}}\}$ be local bases for this
$C^\infty(X)$-module. Then any element $\Phi\in H_{N+2}(\dl_N)$
factorizes
\mar{v160,1}\ben
&& \Phi= \op\sum_{0\leq|\Xi|} \Phi^{r_{N+1},\Xi} d_\Xi
\Delta_{r_{N+1}}\om, \qquad \Phi^{r_{N+1},\Xi}\in
\cS^0_\infty[F;Y], \label{v160}\\
&&\Delta_{r_{N+1}}=G_{r_{N+1}}+ h_{r_{N+1}}=\op\sum_{0\leq|\La|}
\Delta_{r_{N+1}}^{r_N,\La}\ol c_{\La r_N} + h_{r_{N+1}},
\label{v161}
\een
via elements of $\cC_{(N+1)}$. Clearly, this factorization is
independent of specification of local bases
$\{\Delta_{r_{N+1}}\}$. Let us extend the BGDA $\cP^*_\infty\{N\}$
(\ref{v91}) to the  BGDA $\cP^*_\infty\{N+1\}$ possessing local
bases
\be
\{s^A,\ol s_A, \ol c_r, \ol c_{r_1}, \ldots, \ol c_{r_N}, \ol
c_{r_{N+1}}\}, \quad {\rm Ant}[\ol c_{r_{N+1}}]=N+3, \quad [\ol
c_{r_{N+1}}]=([\Delta_{r_{N+1}}]+1){\rm mod}\,2.
\ee
It is provided with the nilpotent graded derivation
\mar{v170}\beq
\dl_{N+1}=\dl_N + \rdr^{r_{N+1}} \Delta_{r_{N+1}} \label{v170}
\eeq
of antifield number -1. With this graded derivation, the module
$\cP^{0,n}_\infty\{N+1\}_{\leq N+4}$ of graded densities of
antifield number Ant$[\f]\leq N+4$ is split into the chain complex
\mar{v171}\ben
&&0\lto \im \ol\dl \llr^{\ol\dl} \cP^{0,n}_\infty[\ol Y^*;F;Y;\ol
F^*]_1\llr^{\dl_0} \cP^{0,n}_\infty\{0\}_2\llr^{\dl_1}
\cP^{0,n}_\infty\{1\}_3\cdots
 \label{v171}\\
&& \quad \llr^{\dl_{N-1}} \cP^{0,n}_\infty\{N-1\}_{N+1}
 \llr^{\dl_N} \cP^{0,n}_\infty\{N\}_{N+2}\llr^{\dl_{N+1}}
\cP^{0,n}_\infty\{N+1\}_{N+3}\llr^{\dl_{N+1}}
\cP^{0,n}_\infty\{N+1\}_{N+4}. \nonumber
\een
It is readily observed that this complex is $(N+2)$-exact. In this
case, the $(N+1)$-stage Noether identities (\ref{v145}) come from
the complete $(N+1)$-stage Noether identities
\be
 \op\sum_{0\leq|\La|} \Delta_{r_{N+1}}^{r_N,\La} d_\La \Delta_r\om +\ol\dl
h_{r_{N+1}}\om =0, \label{v162}
\ee
which are reproduced as the nilpotency conditions of the graded
derivation (\ref{v170}).
\end{proof}

The iteration procedure based on Theorem \ref{v163} can be
prolonged up to an integer $N_{\rm max}$ when the $N_{\rm
max}$-stage Noether identities are irreducible, i.e., the homology
$H_{N_{\rm max}+2}(\dl_{N_{\rm max}})$ is trivial. This iteration
procedure may also be infinite. It results in the manifested exact
Koszul--Tate complex with the Koszul--Tate boundary operator whose
nilpotency conditions reproduce all Noether and higher Noether
identities of an original Lagrangian system.

\bigskip
\bigskip

\noindent {\bf V. EXAMPLE}
\bigskip

Let us consider a fiber bundle
\mar{v180}\beq
Y=\Bbb R\op\times_X \op\w^{n-1} T^*X, \label{v180}
\eeq
coordinated by $(x^\la, A, B_{\m_1\ldots \m_{n-1}})$. The
corresponding BGDA is $\cS^*_\infty[Y]=\cO^*_\infty Y$. There is
the canonical $(n-1)$-form
\be
B=\frac{1}{(n+1)!}B_{\m_1\ldots \m_{n-1}}dx^{\m_1}\w\cdots\w
dx^{\m_{n-1}} \in \cO^*_\infty Y
\ee
on $Y$ (\ref{v180}). A Lagrangian of topological BF theory in
question reads
\mar{v182}\beq
L_{\rm BF}=\frac1n Ad_HB. \label{v182}
\eeq
The corresponding Euler--Lagrange operator (\ref{0709}) takes the
form
\mar{v183}\ben
&&\dl L= dA\w \cE\om + dB_{\m_1\ldots \m_{n-1}}\w \cE^{\m_1\ldots
\m_{n-1}}\om \nonumber\\
&& \cE=\e^{\m\m_1\ldots \m_{n-1}} d_\m B_{\m_1\ldots \m_{n-1}},
\qquad \cE^{\m_1\ldots \m_{n-1}} = - \e^{\m\m_1\ldots
\m_{n-1}}d_\m A, \label{v183}
\een
where $\e$ is the Levi--Civita symbol.

Let us extend the BGDA $\cO^*_\infty Y$ to the BGDA
$\cP^*_\infty[\ol Y^*;Y]$ where
\be
VY=Y\op\times_X Y, \qquad \ol Y^*= (\Bbb R\op\times_X
\op\w^{n-1}TX)\op\ot_X \op\w^n T^*X.
\ee
This BGDA  possesses the local bases $\{ A, B_{\m_1\ldots
\m_{n-1}}, \ol s, \ol s^{\m_1\ldots \m_{n-1}}\}$, where $\ol s,
\ol s^{\m_1\ldots \m_{n-1}}$ are odd of antifield number 1. With
the nilpotent Koszul--Tate differential
\be
\ol\dl=\frac{\rdr}{\dr \ol s}\cE + \frac{\rdr}{\dr \ol
s^{\m_1\ldots \m_{n-1}}} \cE^{\m_1\ldots \m_{n-1}},
\ee
we have the complex (\ref{v042}):
\be
0\lto \im\ol\dl \llr^{\ol\dl} \cP^{0,n}_\infty[\ol Y^*;Y]_1
\llr^{\ol\dl} \cP^{0,n}_\infty[\ol Y^*;Y]_2.
\ee
A generic one-chain reads
\be
\Phi= \op\sum_{0\leq |\La|}(\Phi^\La\ol s_\La +
\Phi^\La_{\m_1\ldots \m_{n-1}} \ol s^{\m_1\ldots
\m_{n-1}}_\La)\om,
\ee
and the cycle condition $\ol\dl\Phi=0$ takes the form
\mar{v189}\beq
\Phi^\La\cE_\La + \Phi^\La_{\m_1\ldots \m_{n-1}} \cE^{\m_1\ldots
\m_{n-1}}_\La=0. \label{v189}
\eeq
If $\Phi^\La$ and $\Phi^\La_{\m_1\ldots\m_{n-1}}$ are independent
of the variational derivatives (\ref{v183}) (i.e., $\Phi$ is a
nontrivial cycle), the equality (\ref{v189}) is split into the
following two ones
\mar{v187,8}\ben
&& \Phi^\La\cE_\La=0, \label{v187}\\
&&  \Phi^\La_{\m_1\ldots \m_{n-1}} \cE^{\m_1\ldots
\m_{n-1}}_\La=0. \label{v188}
\een
The equality (\ref{v187}) holds iff $\Phi^\La=0$, i.e., there is
no Noether identities for $\cE$. The equality (\ref{v188}) is
satisfied iff
\be
\Phi^{\la_1\ldots \la_k}_{\m_1\ldots\m_{n-1}}\e^{\m\m_1\ldots
\m_{n-1}}=- \Phi^{\m\la_2\ldots
\la_k}_{\m_1\ldots\m_{n-1}}\e^{\la_1\m_1\ldots \m_{n-1}}.
\ee
It follows that $\Phi$ factorizes as
\be
\Phi= \op\sum_{0\leq |\Xi|} G_{\nu_2\ldots\nu_{n-1}}^\Xi
d_\Xi\Delta^{\nu_2\ldots\nu_{n-1}}\om
\ee
via local graded densities
\mar{v190}\beq
\Delta^{\nu_2\ldots\nu_{n-1}}=\Delta^{\nu_2\ldots\nu_{n-1},
\la}_{\al_1\ldots\al_{n-1}}\ol
s^{\al_1\ldots\al_{n-1}}_\la=\dl^\la_{\al_1}\dl^{\nu_2}_{\al_2}\cdots
\dl^{\nu_{n-1}}_{\al_{n-1}} \ol s^{\al_1\ldots\al_{n-1}}_\la=
d_{\nu_1}\ol s^{\nu_1\nu_2\ldots\nu_{n-1}}, \label{v190}
\eeq
which provide the complete Noether identities$^1$
\mar{v191}\beq
d_{\nu_1}\cE^{\nu_1\nu_2\ldots\nu_{n-1}}=0. \label{v191}
\eeq

The local graded densities (\ref{v190}) form the bases of a
projective $C^\infty(X)$-module of finite rank which is isomorphic
to the module of sections of the vector bundle
\be
\ol V^*=\op\w^{n-2} TX\op\ot_X \op\w^n T^*X, \qquad V= \op\w^{n-2}
T^*X.
\ee
Therefore, let us extend the BGDA $\cP^*_\infty[\ol Y^*;Y]$ to the
BGDA $\cP^*_\infty\{0\}= \cP^*_\infty[\ol Y^*;Y;V]$ possessing the
local bases
\be
\{A, B_{\m_1\ldots \m_{n-1}}, \ol s, \ol s^{\m_1\ldots \m_{n-1}},
\ol c^{\m_2\ldots \m_{n-1}}\},
\ee
 where $\ol
c^{\m_2\ldots \m_{n-1}}$ are even of antifield number 2. Let
\be
\dl_0= \ol\dl + \frac{\rdr}{\dr \ol c^{\m_2\ldots \m_{n-1}}}
\Delta^{\m_2\ldots \m_{n-1}}
\ee
be its nilpotent graded derivation. Its nilpotency is equivalent
to the Noether identities (\ref{v191}). Then have the one-exact
complex
\be
0\lto \im\ol\dl \llr^{\ol\dl} \cP^{0,n}_\infty[\ol Y^*;Y]_1
\llr^{\dl_0} \cP^{0,n}_\infty\{0\}_2 \llr^{\dl_0}
\cP^{0,n}_\infty\{0\}_3,
\ee
and so on. Iterating the arguments we come to the following
$(N+1)$-exact complex (\ref{v94}) for $N\leq n-3$.

Let us consider the vector bundles
\be
V_k=\op\w^{n-k-2} T^*X, \qquad k=1,\ldots, N,
\ee
and the corresponding BGDA
\be
\cP^*_\infty\{N\}= \cP^*_\infty[...V_3V_1\ol Y^*;Y;VV_2V_4...],
\ee
possessing the local bases
\be
&&\{A, B_{\m_1\ldots \m_{n-1}}, \ol s, \ol s^{\m_1\ldots
\m_{n-1}}, \ol c^{\m_2\ldots \m_{n-1}},\ldots,\ol c^{\m_{N+2}\ldots \m_{n-1}}\},\\
&& [\ol c^{\m_{k+2}\ldots \m_{n-1}}]=(k+1){\rm mod}\,2, \qquad
{\rm Ant}[\ol c^{\m_{k+2}\ldots \m_{n-1}}]=k+3.
\ee
It is provided with the nilpotent graded derivation
\mar{va202}\ben
&& \dl_N=\dl_0 + \op\sum_{1\leq k\leq N}\frac{\rdr}{\dr \ol
c^{\m_{k+2}\ldots \m_{n-1}}}
 \Delta^{\m_{k+2}\ldots \m_{n-1}}, \nonumber\\
&&  \Delta^{\m_{k+2}\ldots
\m_{n-1}}=d_{\m_{k+1}}c^{\m_{k+1}\m_{k+2}\ldots \m_{n-1}},
\label{va202}
\een
of antifield number -1. The nilpotency results from the Noether
identities (\ref{v191}) and the equalities
\mar{v212}\beq
d_{\m_{k+2}}\Delta^{\m_{k+2}\ldots \m_{n-1}}=0, \qquad,
k=0,\ldots,N, \label{v212}
\eeq
which are $k$-stage Noether identities.$^1$ Then the above
mentioned $(N+1)$-exact complex is
\mar{v203}\ben
&&0\lto \im \ol\dl \llr^{\ol\dl} \cP^{0,n}_\infty[\ol
Y^*;Y]_1\llr^{\dl_0} \cP^{0,n}_\infty\{0\}_2\llr^{\dl_1}
\cP^{0,n}_\infty\{1\}_3\cdots
\label{v203}\\
&& \qquad
 \llr^{\dl_{N-1}} \cP^{0,n}_\infty\{N-1\}_{N+1}
\llr^{\dl_N} \cP^{0,n}_\infty\{N\}_{N+2}\llr^{\dl_N}
\cP^{0,n}_\infty\{N\}_{N+3}. \nonumber
\een
It obeys the $(N+2)$-homology regularity condition as follows.

\begin{lem} \label{v220} \mar{v220}
Any $(N+2)$-cycle $\Phi\in \cP^{0,n}_\infty\{N-1\}_{N+2}$ up to a
$\dl_{N-1}$-boundary takes the form
\mar{v218}\ben
&& \Phi=\op\sum_{k_1+\cdots +k_i+3i=N+2}\sum_{0\leq|\La_1|,\ldots,
|\La_i|}G^{\La_1\cdots \La_i}_{\m^1_{k_1+2}\ldots
\m^1_{n-1};\ldots; \m^i_{k_i+2}\ldots \m^i_{n-1}} \label{v218} \\
&&\qquad d_{\La_1} \Delta^{\m^1_{k_1+2}\ldots \m^1_{n-1}}\cdots
d_{\La_i} \Delta^{\m^i_{k_i+2}\ldots \m^i_{n-1}}\om, \qquad
k=-1,0,1,\ldots, N, \nonumber
\een
where $k=-1$ stands for
\be
\ol c^{\m_1\ldots\m_{n-1}}=\ol s^{\m_1\ldots\m_{n-1}}, \qquad
\Delta^{\m_1\ldots\m_{n-1}}=\cE^{\m_1\ldots\m_{n-1}}.
\ee
It follows that $\Phi$ is a $\dl_N$-boundary.
\end{lem}

\begin{proof}
Let us choose some basis element $\ol c^{\m_{k+2}\ldots \m_{n-1}}$
and denote it simply by $\ol c$. Let $\Phi$ contain a summand
$\f_1 \ol c$, linear in $\ol c$. Then the cycle condition reads
\be
\dl_{N-1}\Phi=\dl_{N-1}(\Phi-\f_1 \ol c) + (-1)^{[\ol
c]}\dl_{N-1}(\f_1)\ol c + \f \Delta=0, \qquad \Delta=\dl_{N-1}\ol
c.
\ee
It follows that $\Phi$ contains a summand $\psi\Delta$ such that
\be
(-1)^{[\ol c]+1}\dl_{N-1}(\psi)\Delta +\f\Delta=0.
\ee
This equality implies the relation
\mar{v213}\beq
\f_1=(-1)^{[\ol c]+1}\dl_{N-1}(\psi) \label{v213}
\eeq
because the reduction conditions (\ref{v212}) involve total
derivatives of $\Delta$, but not $\Delta$. Hence,
\be
\Phi=\Phi' +\dl_{N-1}(\psi \ol c),
\ee
where $\Phi'$ contains no term linear in $\ol c$. Furthermore, let
$\ol c$ be even and $\Phi$ has a summand $\sum \f_r \ol c^r$
polynomial in $\ol c$. Then the cycle condition leads to the
equalities
\be
\f_r\Delta=-\dl_{N-1}\f_{r-1}, \qquad r\geq 2.
\ee
Since $\f_1$ (\ref{v213}) is $\dl_{N-1}$-exact, then $\f_2=0$ and,
consequently, $\f_{r>2}=0$. Thus, a cycle $\Phi$ up to a
$\dl_{N-1}$-boundary contains no term polynomial in $\ol c$. It
reads
\mar{v217}\beq
\Phi=\op\sum_{k_1+\cdots +k_i+3i=N+2}\sum_{0<|\La_1|,\ldots,
|\La_i|}G^{\La_1\cdots \La_i}_{\m^1_{k_1+2}\ldots
\m^1_{n-1};\ldots; \m^i_{k_i+2}\ldots \m^i_{n-1}} \ol
c^{\m^1_{k_1+2}\ldots \m^1_{n-1}}_{\La_1}\cdots \ol
c_{\La_i}^{\m^i_{k_i+2}\ldots \m^i_{n-1}}\om. \label{v217}
\eeq
However, the terms polynomial in $\ol c$ may appear under general
covariant transformations
\be
\ol c'^{\nu_{k+2}\ldots \nu_{n-1}}=\det(\frac{\dr x^\al}{\dr
x'^\bt}) \frac{\dr x'^{\nu_{k+2}}}{\dr x^{\m_{k+2}}}\cdots
\frac{\dr x'^{\nu_{n-1}}}{\dr x^{\m_{n-1}}}\ol c^{\m_{k+2}\ldots
\m_{n-1}}
\ee
of a chain $\Phi$ (\ref{v217}). In particular, $\Phi$ contains the
summand
\be
\op\sum_{k_1+\cdots +k_i+3i=N+2}F_{\nu^1_{k_1+2}\ldots
\nu^1_{n-1};\ldots; \nu^i_{k_i+2}\ldots \nu^i_{n-1}} \ol
c'^{\nu^1_{k_1+2}\ldots \nu^1_{n-1}}\cdots \ol
c'^{\nu^i_{k_i+2}\ldots \nu^i_{n-1}},
\ee
which must vanish if $\Phi$ is a cycle. This takes place only if
$\Phi$ factorizes through the graded densities
$\Delta^{\m_{k+2}\ldots \m_{n-1}}$ (\ref{va202}) in accordance
with the expression (\ref{v218}).
\end{proof}

Following the proof of Lemma \ref{v220}, one can show that any
$(N+2)$-cycle $\Phi\in \cP^{0,n}_\infty\{N\}_{N+2}$ up to a
boundary takes the form
\mar{v221}\beq
\Phi=\op\sum_{0\leq|\La|}G^\La_{\m_{N+2}\ldots \m_{n-1}}
\Delta^{\m_{N+2}\ldots \m_{n-1}}\om, \label{v221}
\eeq
i.e., the homology $H_2(\dl_N)$ of the complex (\ref{v203}) is
finitely generated by the cycles $\Delta^{\m_{N+2}\ldots
\m_{n-1}}$. Thus, the complex (\ref{v203}) admits the
$(N+2)$-exact extension (\ref{v171}).

The iteration procedure is prolonged till $N=n-3$. Given the BGDA
$\cP^*\{n-3\}$, the corresponding $(n-2)$-exact complex
(\ref{v203}) has the following $(n-1)$-exact extension. Let us
consider the BGDA $\cP^*\{n-2\}$, where $V_{n-2}=X\times\Bbb R$.
It possesses the local bases
\be
\{A, B_{\m_1\ldots \m_{n-1}}, \ol s, \ol s^{\m_1\ldots \m_{n-1}},
\ol c^{\m_2\ldots \m_{n-1}},\ldots,\ol c^{\m_{n-1}}, \ol c\},
\ee
where $[\ol c]=(n-1){\rm mod}\,2$ and Ant$[\ol c]=n+1$. It is
provided with the nilpotent graded derivation
\mar{v223}\beq
\dl_{n-2}=\dl_0 + \op\sum_{1\leq k\leq n-3}\frac{\rdr}{\dr \ol
c^{\m_{k+2}\ldots \m_{n-1}}} \Delta^{\m_{k+2}\ldots \m_{n-1}} +
\frac{\rdr}{\dr \ol c}\Delta, \qquad \Delta=d_{\m_{n-1}}\ol
c^{\m_{n-1}} \label{v223}
\eeq
Then the above mentioned $(n-1)$-exact complex is
\mar{v224}\ben
&&0\lto \im \ol\dl \llr^{\ol\dl} \cP^{0,n}_\infty[\ol
Y^*;Y]_1\llr^{\dl_0} \cP^{0,n}_\infty\{0\}_2\llr^{\dl_1}
\cP^{0,n}_\infty\{1\}_3\cdots
\label{v224}\\
&& \qquad
 \llr^{\dl_{n-3}} \cP^{0,n}_\infty\{n-3\}_{n-1}
\llr^{\dl_{n-2}} \cP^{0,n}_\infty\{n-2\}_n\llr^{\dl_{n-2}}
\cP^{0,n}_\infty\{n-2\}_{n+1}. \nonumber
\een
Following the proof of Lemma \ref{v220}, one can show that the
$n$-homology regularity condition is satisfied. Therefore, any
$n$-cycle up to a $\dl_{n-3}$-boundary takes the form
\be
\Phi=\op\sum_{0\leq|\La|}G^\La \ol c_\La.
\ee
The cycle condition reads
\be
\dl_{n-2}\Phi=\op\sum_{0\leq|\La|}G^\La d_\La\Delta=0.
\ee
It follows that $G^\La=0$ and, consequently, $\Phi=0$. Thus, the
$n$-homology of the complex (\ref{v224}) is trivial, and this
complex is exact. It is a desired Koszul--Tate complex of a
Lagrangian system in question. The nilpotency conditions of its
boundary operator (\ref{v223}) restarts all the Noether identities
of this Lagrangian system.

\end{document}